\documentclass[12pt]{iopart}
\usepackage{graphicx}
\begin{document}
\title{How extreme are extreme black holes?}

\author{David Garfinkle}
\address{Dept. of Physics, Oakland University,
Rochester, MI 48309, USA}
\address{and Michigan Center for Theoretical Physics, Randall Laboratory of Physics, University of Michigan, Ann Arbor, MI 48109-1120, USA}
\ead{garfinkl@oakland.edu}


\date{\today}

\begin{abstract}

We examine the properties of nearly extremal black holes produced by gravitational collapse.  It is shown that 
an observer who crosses the black hole horizon at late times rapidly encounters a singularity.

\end{abstract}


\maketitle

\section{Introduction}
\indent

Nonrotating black holes with both mass and charge have two horizons: the outer horizon is the actual event horizon of the black hole, while the inner 
horizon is unstable to small perturbations, and is conjectured to become singular in any realistic situation.\cite{israelpoisson}
For such black holes the charge is always less than or equal to the mass, with the special case of charge equals mass called an extreme
black hole.  In the limit as the charge approaches the mass, the two horizons merge.  Thus, as pointed out by Marolf\cite{marolf} one
might expect that in some sense extreme black holes have singular event horizons.  More precisely, it is claimed in \cite{marolf} that 
an observer crossing the event horizon encounters a singularity a short time later and that this short time goes to zero in the limit of 
extremality.  

The treatment of \cite{marolf} used an eternal black hole, whereas physical black holes are formed by gravitational collapse.  In this paper
we treat the problem of nearly extreme black holes formed by gravitational collapse.  For simplicity, we use collapse of a charged thin shell.
We then consider a free fall observer who falls into such a black hole at late times and calculate the time it takes for him to reach the
singularity.  Methods and results are presented in section 2 and conclusions are presented in section 3.

\section{Methods and Results}

A charged, nonrotating, black hole is described by the Reissner-Nordstrom metric
\begin{equation}
d {s^2} = - F d {t^2} + {F^{-1}} d {r^2} + {r^2} (d {\theta ^2} + {\sin ^2} \theta d {\phi ^2})
\label{rn}
\end{equation}
where the quantity $F$ is given by 
\begin{equation}
F = 1 - {\frac {2m} r} + {\frac {q^2} {r^2}}
\label{F}
\end{equation}
where $m$ and $q$ are respectively the mass and charge of the black hole.  The quantity $F$ vanishes at the outer and inner
horizons of the black hole, which are located respectively at $r=r_+$ and $r=r_-$ where
\begin{equation}
{r_\pm} = m \pm {\sqrt {m^2 - q^2}}
\end{equation}
The coordinate system in eqn.(\ref{rn}) becomes singular at the horizons, thus to describe the process of gravitational collapse
and black hole formation, we need a different coordinate system.  We therefore use Eddington-Finkelstein coordinates 
($v,r,\theta,\phi$) where $v$ is given by
\begin{equation}
v = t + \int {F^{-1}} d r
\end{equation}
In these coordinates the Reissner-Nordstrom metric becomes 
\begin{equation}
d {s^2} = - F d {v^2} + 2 dv \, dr + {r^2} (d {\theta ^2} + {\sin ^2} \theta d {\phi ^2})
\label{rnef}
\end{equation}

\begin{figure}
\includegraphics{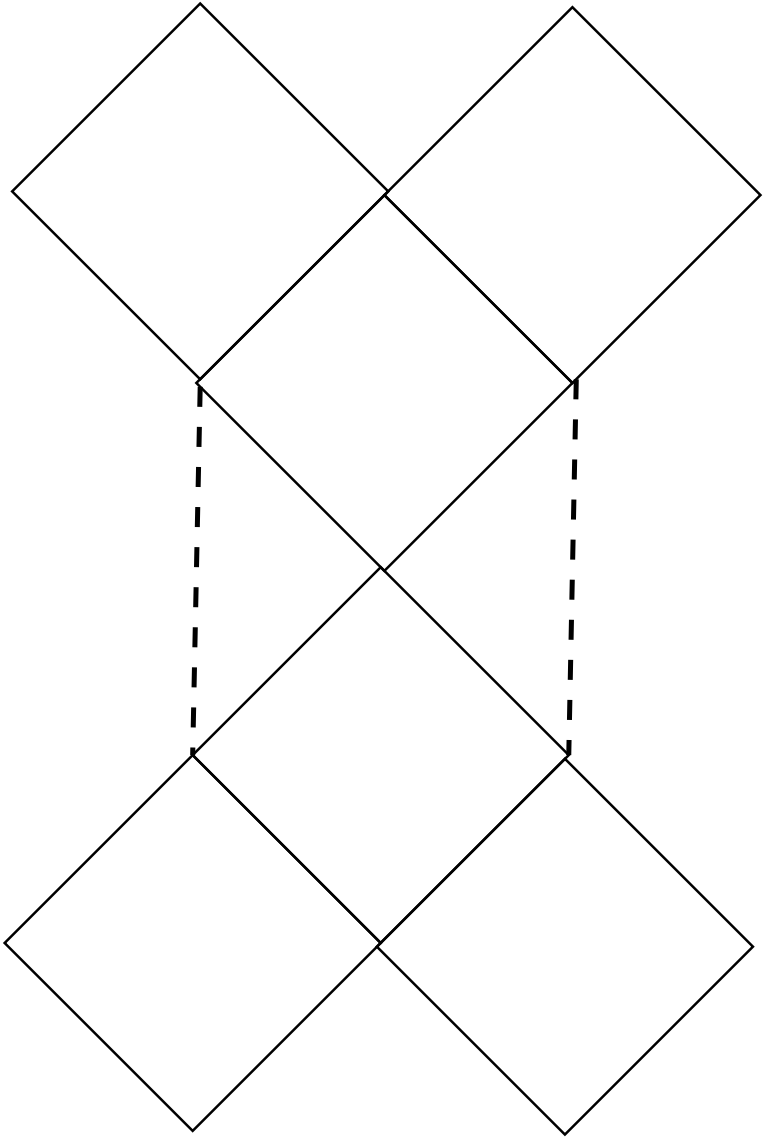}
\caption{\label{fig1} The extended Reissner-Nordstrom spacetime.  Here the outer and inner horizons are 
represented by solid lines at 45 degree angles to the vertical, while the timelike singularity is
denoted by a dashed line }
\end{figure}

The coordinate $v$ is constant along ingoing light rays.  One can also define an analogous coordinate $u$ which is constant
along outgoing light rays.  With a series of coordinate patches of this type, one can cover the extended Reissner-Nordstrom spacetime,
which is shown in fig. (\ref{fig1}).  However, the extended Reissner-Nordstrom spacetime is too large to represent what actually
happens in gravitational collapse.  This is for two different reasons: first when a charged, spherical object collapses, it is only
the spacetime outside the object that is described by the Reissner-Nordstrom metric.  Thus, to describe gravitational collapse, one
must find the path of the outer surface of the collapsing object, and then cut out all parts of the diagram of fig. (\ref{fig1})
that are outside that surface and replace them with something appropriate for the interior of the collapsing object.  The second
reason is that small perturbations of the Reissner-Nordstrom solution blow up as $v \to \infty$ (which is also a null surface at
$r={r_-}$).  Therefore, to get the appropriate diagram, one must replace this null surface with a null singularity and cut out all
portions of the diagram to the future of this surface.  In the end one obtains a restricted portion of the extended Reissner-Nordstrom
spacetime, which can be covered by a single coordinate patch of the sort used in eqn. (\ref{rnef}).   

We want a simple model for gravitational collapse to form the black hole, so we choose a thin shell of charged dust.  
Thin shells in general relativity are described by the Israel formalism\cite{israel} 
in which the history of the shell is a timelike hypersurface $\Sigma$
separating two spacetimes (labeled $+$ and $-$ and which are respectively the exterior and the interior of the shell).  
The intrinsic metric $h_{ab}$ is continuous, while the extrinsic curvature $K_{ab}$ is discontinuous and satisfies  
\begin{equation}
[{K_{ab}}]  = - 8 \pi  \left ( {t_{ab}} - {\textstyle {\frac 1 2}} t {h_{ab}} \right )
\label{israelshell}
\end{equation}
where $t_{ab}$ is the delta function stress-energy of the shell.
Here, for any quantity $\Phi$ we define $[\Phi ]= {\Phi _+}-{\Phi _-}$ where the quantity $\Phi_+$ is $\Phi$ on 
$\Sigma$ evaluated from the $+$ side (and correspondingly for $\Phi _-$).   

The worldsheet of the shell is given by $r=R(\tau)$ where $\tau $ is the proper time of an observer in the shell.  
The stress-energy of the dust shell takes the form
\begin{equation}
{t_{ab}} = \sigma {u_a} {u_b}
\label{shellstress}
\end{equation}
where $u^a$ is the four-velocity of the shell and $\sigma$ is the shell's mass per unit area.  
As shown by de la Cruz and Israel\cite{israelrn}, Kuchar\cite{kuchar} and Chase\cite{chase} eqn.(\ref{israelshell})
can be used to find the equation of motion of the shell as follows: the exterior of the shell is a Reissner-Nordstrom spacetime.
Spherical symmetry implies that the extrinsic curvature takes the form
\begin{equation}
{K_{ab}} = B {u_a}{u_b} + C ({h_{ab}} + {u_a}{u_b})
\label{Kdecomp}
\end{equation}  
Combining eqns. (\ref{israelshell}), (\ref{shellstress}) and (\ref{Kdecomp}) yields
\begin{equation}
[B] = - 4 \pi \sigma
\label{Bjump}
\end{equation}  
\begin{equation}
[C] = - 4 \pi \sigma
\label{Cjump}
\end{equation}
A straightforward calculation using eqn.(\ref{rnef}) shows that the quantities $B$ and $C$ are given by 
\begin{eqnarray}
B = - {{({{\dot R}^2} + F)}^{-1/2}} \left ( {\ddot R} + {\textstyle {\frac 1 2}} {\frac {dF} {dR}}\right )
\label{Beqn}
\\
C = {\frac 1 R} {{({{\dot R}^2} + F)}^{1/2}}
\label{Ceqn}
\end{eqnarray}
where an overdot denotes derivative with respect to $\tau$.
Since the interior of the shell is Minkowski spacetime, expressions for $B$ and $C$ for the shell interior are 
given by eqns (\ref{Beqn}) and (\ref{Ceqn}) with $F$ replaced by $1$.  Applying eqn. (\ref{Ceqn}) to eqn. (\ref{Cjump})
yields 
\begin{equation}
{{({{\dot R}^2} + 1)}^{1/2}} \; - \; {{({{\dot R}^2} + F)}^{1/2}} = 4 \pi R \sigma
\label{motion1}
\end{equation}  
Then using eqns. (\ref{Bjump}), (\ref{Beqn}), and (\ref{motion1}) yields
\begin{equation}
{\frac d {d\tau}} ( {R^2} \sigma ) = 0
\end{equation}
which means that there is a constant $M$ such that $4 \pi \sigma = M/{R^2}$  Using this result in eqn. (\ref{motion1})
and solving for ${\dot R}^2$ yields
\begin{equation}
{{\dot R}^2} = - 1 + {{\left ( {\frac m M} \; + \; {\frac {{M^2} - {q^2}} {2 M R}} \right ) }^2}
\label{motion2}
\end{equation}

\begin{figure}
\includegraphics{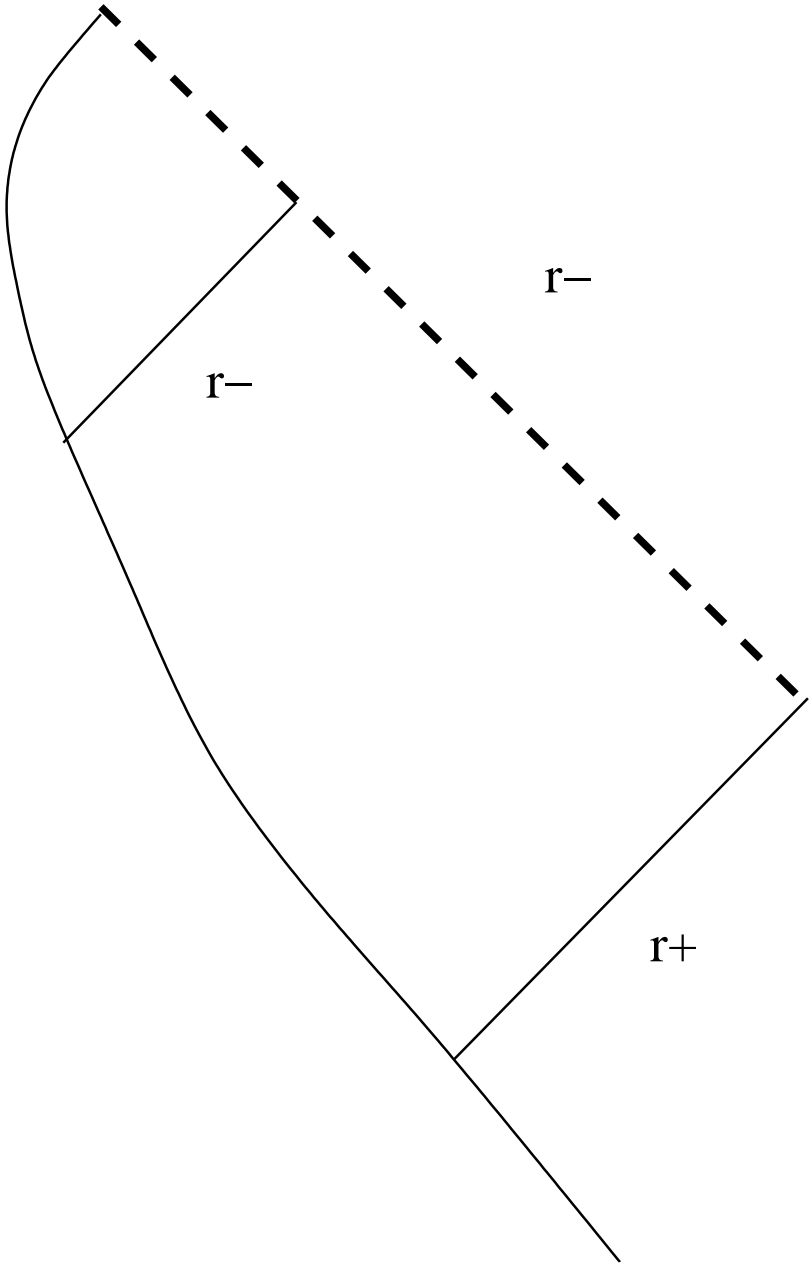}
\caption{\label{fig2} path of the thin shell in the Reissner-Nordstrom spacetime.  Here nonsingular horizons
are denoted by solid lines, and the null singularity by a dashed line }
\end{figure}

We start the shell at a large radius with a speed $c_0$ in the inward direction.  Then it follows from eqn. (\ref{motion2})
that 
\begin{equation}
M = {\frac m {\sqrt {1+ {c_0 ^2}}} } 
\end{equation}
We also introduce the quantity $\epsilon$ defined by
\begin{equation}
\epsilon = {\sqrt {1- {\frac {q^2} {m^2}}}}
\end{equation}
Since we are considering nearly extremal black holes, $\epsilon \ll 1$.  Expressing eqn. (\ref{motion2}) in terms of 
$m, \, \epsilon$ and $c_0$ we obtain
\begin{equation}
{{\dot R}^2} = - 1 + (1 + {c_0 ^2}) {{\left ( 1 + {\frac m {2R}} \left [ {\epsilon ^2} - {\frac {c_0 ^2} {1 + {c_0 ^2}}}
\right ] \right ) }^2}
\label{motion3}
\end{equation} 

We have ${r_\pm} = m (1 \pm \epsilon )$, so $r_-$ is just slightly less than $r_+$ 
and therefore the shell will cross $r_-$ a short (proper) time
after crossing $r_+$.  It is also true that the perturbed black hole becomes singular at $r_-$.  However, it is here that 
the use of Eddington-Finkelstein coordinates becomes essential.  In the collapse process, there are two different null surfaces 
at $r={r_-}$, an outgoing null surface where $v$ is finite, and which is not singular, and an ingoing null surface which
occurs in the limit as $v \to \infty$, and which is singular.  The question then becomes, which of these surfaces does the 
shell cross first?  To answer this question, we note that there is a function $V(\tau )$ such that on the worldsheet of the shell 
$v=V(\tau )$.  
Then from eqn. (\ref{rnef}) it follows that
\begin{equation}
- 1 = - F {{\dot V}^2} + 2 {\dot V} {\dot R}
\end{equation}
and therefore that
\begin{equation}
{\dot V} = {\frac 1 F} \left ( {\dot R} \pm {\sqrt {{{\dot R}^2} + F}}\right )
\end{equation}
However, the null coordinate $v$ always increases towards the future, fixing the sign above to be plus, so we
obtain
\begin{equation}
{\dot V} = {\frac 1 F} \left ( {\dot R} + {\sqrt {{{\dot R}^2} + F}}\right )
\label{dotv}
\end{equation}
Note that in this equation, if ${\dot R} < 0 $ when $F$ vanishes, then $\dot V$ remains finite; but if 
${\dot R} > 0 $ when $F$ vanishes then ${\dot V} \to \infty$.  Since the shell starts out ingoing, and since
it cannot change direction between the two horizons (where $r$ is a timelike coordinate) then it follows that
the shell first crosses $r_-$ at a place where $v$ remains finite, and where therefore the spacetime remains
nonsingular.  After crossing $r_ -$, the shell reaches a minimum radius $r_{\rm min}$ and then re-expands.  It 
follows from eqn. (\ref{motion3}) that
\begin{equation}
{r_{\rm min}} = {\frac m 2} \left ( 1 + {\frac 1 {\sqrt {1 + {c_0 ^2}}}} \right )
\end{equation}    
where we have neglected any terms higher than zeroth order in $\epsilon$.  After reaching $r_{\rm min}$ the
shell re-expands, and eventually reaches $r_-$.  Note, however, that this time the shell approaches $r_-$ with
${\dot R} > 0$.  It then follows from eqn. (\ref{dotv})  that $v$ diverges as $r \to {r_-}$ and thus that 
the shell approaches the null singularity.  
The path of the shell in the usual Penrose diagram of the Reissner-Nordstrom spacetime is shown in Fig. (\ref{fig2}).

The shell clearly takes a non-neglible amount of proper time to encounter the singularity, and this will also be true of an
observer who falls into the black hole shortly after it forms.  However, as we will see, things are somewhat different for
an observer who falls in at late times.  Since for such an observer, $v \gg m$ when the outer horizon is crossed, and since $v$ increases along future directed timelike curves, it follows that $v \gg m$ when the observer encounters the shell.  
Therefore, we are led to consider the behavior of the shell at large $v$.  We know that
as $v \to \infty$ we have $r \to {r_-}$ and therefore $F \to 0$.  However, we will need to know how rapidly $F \to 0$ at large
$v$.  From eqn. (\ref{dotv}) it follows that on the shell we have
\begin{equation}
{\frac {dv} {dr}} = {\frac 1 F} \left ( 1 + {\sqrt { 1 + {\frac F {{\dot R}^2}}}} \right )  
\label{dvdr}
\end{equation}  
Then using eqns. (\ref{F}) and (\ref{dvdr}) and neglecting any terms that are higher order in the small quantities
$F$ and $\epsilon$ we find that on the shell at large $v$ we have
\begin{equation}
{\frac {dv} {dF}} = {\frac {-m} {F {\sqrt {{\epsilon ^2} + F}}}}
\end{equation}
The solution of this equation is 
\begin{equation}
F = {\frac {\epsilon ^2} {{\sinh ^2} \left ( {\frac {\epsilon (v + {c_1})} {2m}}\right ) }}
\label{Flargev}
\end{equation}
where $c_1$ is a constant.  Note that there are two separate regimes where this equation simplifies.  For
$m \ll v \ll m/\epsilon$ we have 
$ F \approx 4 {m^2}/{{(v+{c_1})}^2}$ while for 
$ v \gg m/\epsilon$ we have
$ F \approx 4 {\epsilon ^2} \exp [ -  \epsilon (v+{c_1})/m]$.  

We now consider the fate of a free-fall observer who falls into the black hole at late times.  Just as the shell has
four-velocity ${u^a}=({\dot V},{\dot R})$, denote the four-velocity of the observer by ${v^a} =({\dot v},{\dot r})$.  
Due to the time translation symmetry of the Reissner-Nordstrom spacetime, the observer has a conserved energy per unit 
mass $E$ given by
\begin{equation}
E = F {\dot v} - {\dot r}
\label{energy}
\end{equation}
Since $v^a$ is a unit timelike vector, the same reasoning that led to eqn. (\ref{dotv}) yields
\begin{equation}
{\dot v} = {\frac 1 F} \left ( {\dot r} + {\sqrt {{{\dot r}^2} + F}}\right )
\label{dotv2}
\end{equation}
Using eqns. (\ref{energy}) and (\ref{dotv2}) we find
\begin{equation}
{\dot r} = - {\sqrt {{E^2} - F}}
\end{equation}
At late times, the shell is approximately at $r_-$ and for the region in between $r_+$ and $r_-$ we have
${\dot r} \approx - E$. Thus the observer crosses the shell approximately a proper time $2m\epsilon /E$
after crossing the outer horizon.  

To find the amount of proper time that it takes for the observer
to encounter the singularity, we must also find the behavior of the observer in the interior of the shell.  
To begin with, we calculate the quantity ${v^a}{u_a}$ when the observer crosses the shell.  From eqn. (\ref{rn})
we find
\begin{equation}
{v^a}{u_a} = - F {\dot V}{\dot v} + {\dot V}{\dot r} + {\dot v}{\dot R}
\label{vdotu}
\end{equation}  
Then using eqns. (\ref{dotv}), (\ref{dotv2}) and (\ref{vdotu}) we obtain
\begin{equation}
{v^a}{u_a} = {F^{-1}} \left ( {\dot R}{\dot r} - {\sqrt {{{\dot R}^2} + F}} {\sqrt {{{\dot r}^2} + F}} \right )
\label{vdotu2}
\end{equation} 
Since when the observer crosses the shell we have ${\dot R} > 0$ and ${\dot r} < 0$ it follows that to lowest order
in $\epsilon$
\begin{equation}
{v^a}{u_a} = - 2  {F^{-1}} | {\dot R}{\dot r}|
\label{vdotu3}
\end{equation}
and then using eqn. (\ref{motion3}) we obtain
\begin{equation}
{v^a}{u_a} = -  {F^{-1}} {\frac {E {c_0 ^2}} {\sqrt {1+{c_0 ^2}}}}
\label{vdotu4}
\end{equation}

\begin{figure}
\includegraphics{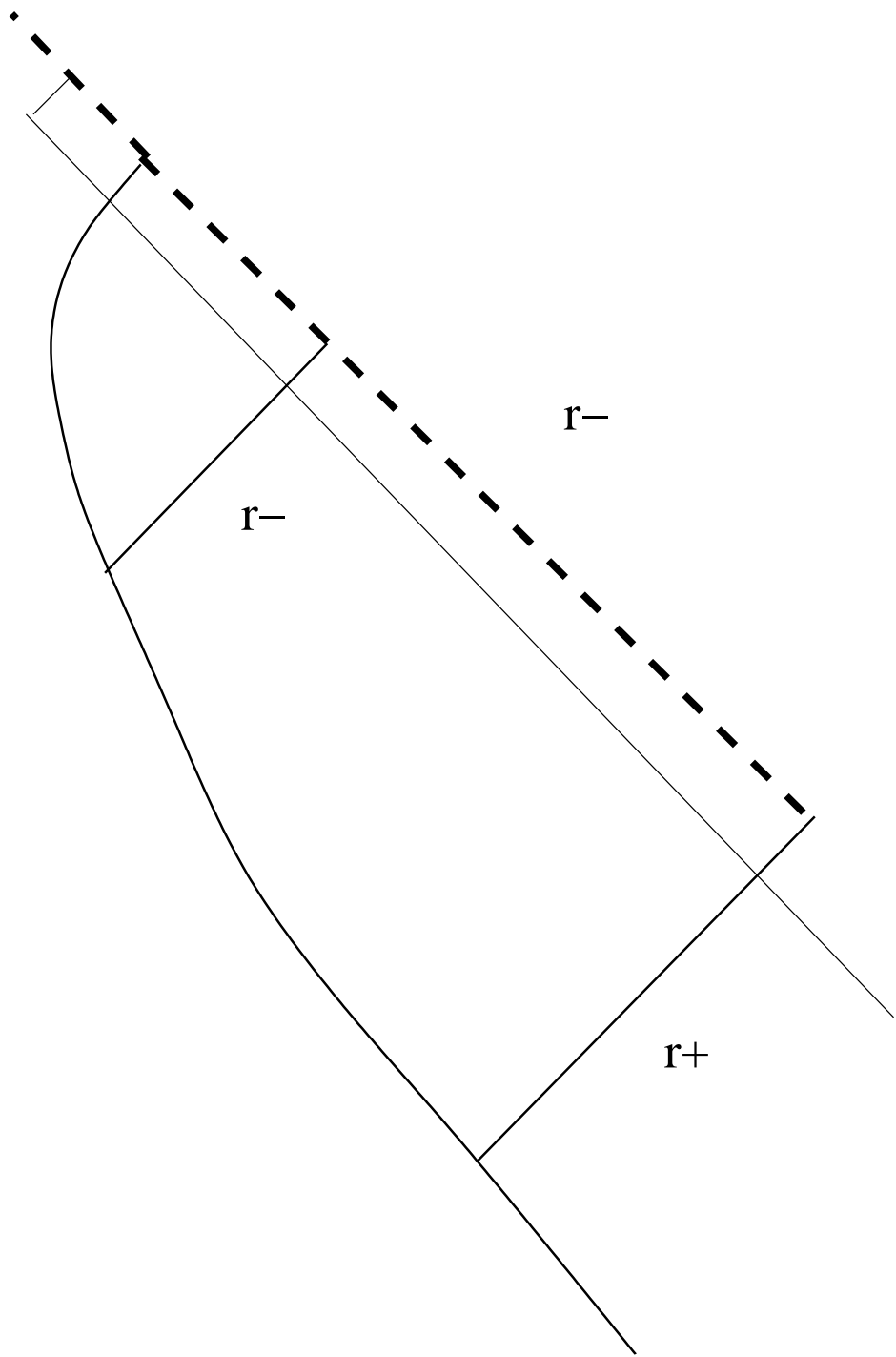}
\caption{\label{fig3} path of the geodesic observer (thin curve) in the spacetime formed by the collapse of the
thin shell.  Here nonsingular horizons are denoted by solid lines, and the null singularity by a dashed line }
\end{figure}

Since $F$ is very small when the observer crosses the shell, it follows that ${v^a}{u_a}$ is very large there.  
However, since $v^a$ satisfies the geodesic equation, and since the Christoffel symbols in the thin shell spacetimes
have only jumps rather than delta functions, it follows that $v^a$ is continuous across the shell.  Since the thin
shell formalism also requires that the metric and $u^a$ are continuous across the shell, it follows that immediately upon
entering the interior of the shell ${v^a}{u_a}$ is still given by the expression in eqn. (\ref{vdotu4}).  But the interior
of the shell is Minkowski spacetime, and the shell has 
\begin{equation}
{\dot R} = {\frac {c_0 ^2} {2 {\sqrt {1 + {c_0 ^2}}}}}
\end{equation}  
So it follows from eqn. (\ref{vdotu4}) that in the rest frame of the center of the shell, the observer is an ingoing
geodesic with 
\begin{equation}
\gamma = {F^{-1}} {\frac {E {c_0 ^2}} {1 + {c_0 ^2}}}
\end{equation}
This is a very large $\gamma$ factor, so even though the observer must travel a distance of $2m$ (to the center and back) 
to encounter the singularity, this only takes a proper time of 
\begin{equation}
\Delta \tau = 2 m F {\frac  {1 + {c_0 ^2}} {E {c_0 ^2}}}
\end{equation}
Since $F$ is given by eqn. (\ref{Flargev}) it follows that this is a very small proper time.  The motion of the observer
is shown in fig. (\ref{fig3})  

\section{Conclusion}

Though the pictures drawn are somewhat different, our calculations agree with the main conclusion of \cite{marolf}.  
For a nearly extremal black hole formed by gravitational collapse an observer crossing the event horizon at late times
almost immediately encounters a curvature singularity.  Note however, that this statement involves two different small 
numbers: $\epsilon$ which expresses how close to extremality the black hole is, and $m/{v_h}$ which measures how late the 
observer crosses the black hole event horizon.  (Here $v_h$ is the value of $v$ at which the observer crosses the horizon).
In all cases, the proper time from event horizon to shell is of order $m \epsilon$.  
For $m/{v_h} \gg \epsilon$ the proper time from shell to singularity is of order ${m^3}/{v_h ^2}$, while for   
$m/{v_h} \ll \epsilon$ the proper time from shell to singularity is completely negligible.  

We now consider the effect the nature of the singularity could have on this result.  Throughout, we have assumed that the singularity
is null.  However, it is generally thought that when black holes form the singularity starts out null near the horizon and 
then becomes spacelike further inside the black hole.  However, any such change would only serve to decrease the amount of time
it would take for an observer to encounter the singularity, so the conclusion that the observer encounters the singularity almost
immediately after crossing the event horizon remains true.

Finally note that the collapsing thin shell is only one simple model for the formation of a charged black hole by gravitational collapse.  It would be interesting to consider other models ({\it e.g.} the collapse of a charged scalar field) to see whether
the features found in the shell model continue to hold.

\ack
I would like to thank Ratin Akhoury, Ryo Saotome, Eric Poisson, Gary Horowitz, and Don Marolf for useful discussions. 
This work was supported by NSF grant PHY-0855532 to Oakland University.

\end{document}